# Investigation of Mixed Irradiation Effects in p-MCz Thin Silicon Microstrip Detector for the HL-LHC Experiments

**Shilpa Patyal[a], Nitu Saini[a], Balwinder Kaur[a], Puspita Chatterjee[a], Ajay K. Srivastava[a,1]**

[a]*Department of Physics, University Institute of Sciences, Chandigarh University, Gharuan - Mohali, Punjab, 140413, India.*
*E-mail:* kumar.uis@cumail.in

ABSTRACT: A lot of R&D work is carried out in the CERN RD 50 collaboration to find out the best material for the Si detectors that can be used in the harsh radiation environment of HL- LHC, n and p-MCz Si was identified as one of the prime candidates as a material for strip detector that can be chosen for the phase 2 upgrade plan of the new Compact Muon Solenoid Tracker detector in 2026.

In this work, four level deep-trap mixed irradiation model for p-MCz Si is proposed by the comparison of experimental data on the full depletion voltage and leakage current to the Shockley Read Hall recombination statistics results on the mixed irradiated p-MCz Si PAD detector. The effective introduction rate ($\eta_{eff}$) of shallower donor deep trap E30K is extracted using SRH theory calculations for experimental $N_{eff}$ and that can show the behavior of space charges and electric field distribution in the p-MCz Si strip detector and compared its value with the $\eta_{eff}$ of shallower donor deep trap E30K in then MCz Si microstrip detector.

Prediction uncertainty in the p-MCz Si radiation damage mixed irradiation model is considered in the full depletion voltage and leakage current. A very good agreement is observed in the experimental and SRH results. This radiation damage model is also used to extrapolate the value of the full depletion voltage at different mixed (proton + neutron) higher irradiation fluences for the thin p-MCz Si microstrip detector.

---

[1] Corresponding author. Kumar.uis@cumail.in

**Contents**

## 1. Introduction

Within CERN RD50 European collaboration, n in p Si detectors shows good performance for the new CMS Tracker at HL-LHC experiments [1-2]. The p-MCz (Magnetic Czochralski) Si can be a suitable candidate as a material for n in p Si strip detector. Therefore, it is mandatory to investigate the radiation damage effects in the thin n in p-MCz Si microstrip detector [3-6]. The macroscopic and microscopic performance of these Si strip detectors have been measured using current-voltage (I/V), capacitance-voltage (C/V), Thermally Stimulated Current (TSC), Deep Level Transient spectroscopy (DLTS), Transient Current Technique (TCT) and Alibava system SL, Barcelona, Spain setup.

    The paper describes the following sections; section 2 shows the four level deep trap model, detector model and Schockley Read Hall (SRH) calculations for the E (30K) and section3 elaborates the results and discussion on the extrapolated values on the full depletion voltage and leakage current at higher mixed irradiated fluences.

## 2. Four level deep traps mixed irradiation damage model for p-MCz Si PAD detector

In this section, four level deep-traps mixed irradiation model for p-MCz Si PAD detector is proposed for the study of bulk radiation damage effects on $V_{fd}$ and leakage current.

**2.1 Si PAD Detector model and SRH/ TCAD device calculations**

A rectangular cell of 0.0625 cm$^2$ x 300 μm p-MCz Si PAD detector model is used for the SRH calculations (see figure 1) and the device and process parameters of the detector are shown in (Table 1). The effective doping concentration ($N_{eff}$) and generation leakage current ($I_{gen}$) at full depletion voltage ($V_{fd}$) can be calculated by the SRH theoretical expressions[5] in p-MCz Si PAD detector irradiated by mixed neutron and proton fluences[3] (1 MeV equivalent neutron) 3.13x10$^{14}$, 4.98 x 10$^{14}$, 8.82 x 10$^{14}$ n$_{eq.}$/cm$^{-2}$, where symbols are having their usual meaning.

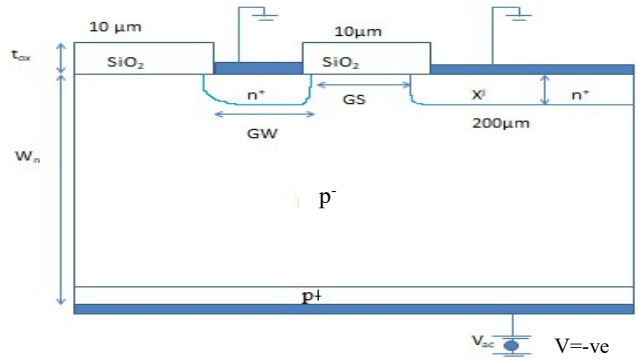

**Figure 1**. Cross-section of the 0.0625 cm² x 300μm n⁺ p-MCz Si pad detector model used in the present study for SRH calculations.

Table 1. Device and process parameters of p-MCz Si PAD detector for mixed irradiation.

| S.No. | Physical parameters | Values |
|---|---|---|
| 1. | Doping concentration ($N_A$) | $2.87 \times 10^{12}$ cm⁻³ |
| 2. | Oxide thickness ($t_{ox}$) | 0.5 μm |
| 3. | Junction Depth ($X_j$) | 1 μm |
| 4. | Guard ring spacing (GS) | 10 μm |
| 5. | Guard ring width (GW) | 100 μm |
| 6. | Device depth ($W_n$) | 300 μm |
| 7. | Fixed oxide charge ($Q_f$) | $1 \times 10^{12}$ cm⁻² |

## 2.1.1 Comparison of experimental and theoretical $V_{fd}$ (SRH) / TCAD device simulations

To verify the experimental data on $V_{fd}$ and leakage current in mixed irradiated p-MCZ Si PAD detector, the microscopic parameters are fed into SRH / TCAD for the comparison of the experimental data and theoretical $V_{fd}$ (SRH) / TCAD device simulation.

A few of microscopic parameters (E5, H(152K), E(30K) are necessary to tune for good fit of the experimental results on $V_{fd}$ and leakage current to our SRH / TCAD simulation calculations. As a result, four – level deep traps model for mixed irradiation is proposed, which is shown in Table 2, where E5,$C_iO_i$,E(30K),H(152K) are the four deep trap defects. E5 and H(152K) are the acceptor deep level traps here, E5 is mainly responsible for the increase of leakage current in mixed irradiated detectors while H(152K) is responsible for the increase of the negative space charge with increasing fluences. $C_iO_i$ and E (30K) are donor trap defects and these defects increases the positive space charge and it has been observed that the effective introduction rate of E(30K) in n-MCz or in p-MCz Si detector irradiated by proton, neutron or mixed plays an important role and that can be a key trap to explain the macroscopic $V_{fd}$ performance of the n / p-MCz PAD detector compensates partially the increase of the negative space charge introduced by the defect H(152K).

Table 2. Mixed irradiation "four level deep traps model" for MCz-p Si PAD detector.

| Defect/type | Effects on the macroscopic parameters | Energy level(eV) | $\sigma_n$ (capture crossection of electrons) [cm²] | $\sigma_p$ (capture crossection of holes) [cm²] | η ( introduction rate)[cm⁻¹] |
|---|---|---|---|---|---|
| E5/Acceptor | Increase of leakage current | Ec-0.46eV | $1.41\times10^{-15}$ | $2.79\times10^{-15}$ | 12.4 |
| H(152K)/Acceptor | -ve space charge | Ev+0.42 eV | $4.58\times10^{-13}$ | $6.15\times10^{-13}$ | 0.04 |
| $C_iO_i$/Donor | +ve space charge | EV+0.36 eV | $2.08\times10^{-18}$ | $2.45\times10^{-15}$ | 1.1 |
| E(30K)/Donor | +ve space charge | Ec-0.10eV | $2.30\times10^{-14}$ | $2.00\times10^{-15}$ | See fig.2(a) |

The Experimental and SRH values of $V_{fd}$ for the 300μm thick p-MCz PAD detectors are shown a good agreement (see Figure 2(a)), whereas, the leakage current is also in good agreement with 10% uncertainty in the experiment value of the leakage current (see Figure2(b)).

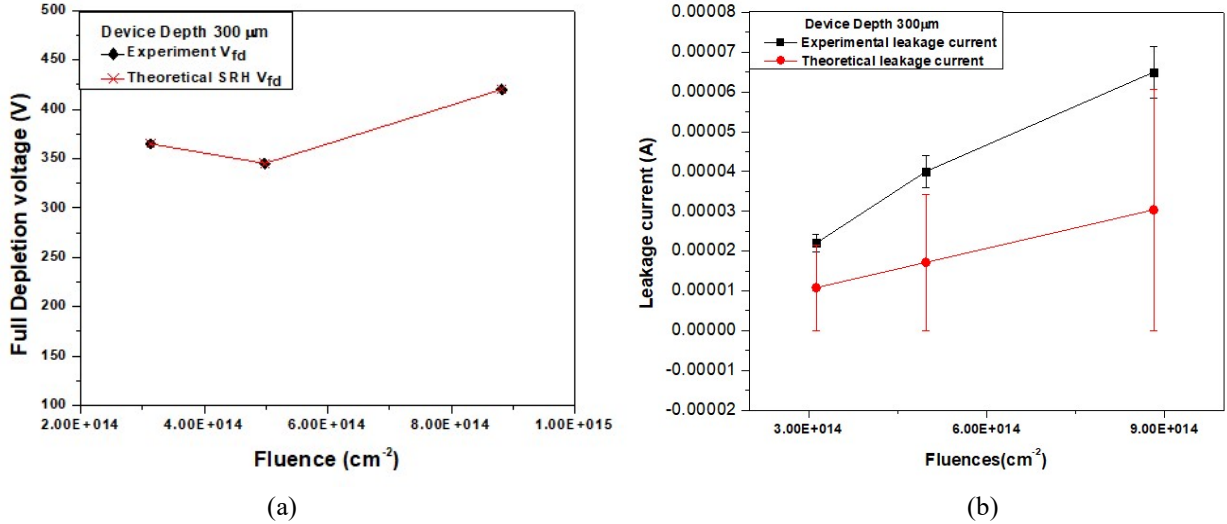

(a) (b)

**Figure 2.**(a)Comparison of Experimental and Theoretical SRH value of full depletion voltage ($V_{fd}$) in mixed irradiated p-MCz Si PAD detector. (b) Leakage current as a function of mixed irradiated fluences at 297K after beneficial annealing of 80 min 60 degree.

Figure 3 shows the full depletion voltage as a function of mixed irradiation fluences in a 200 μm detector in p-MCz Si PAD detector.

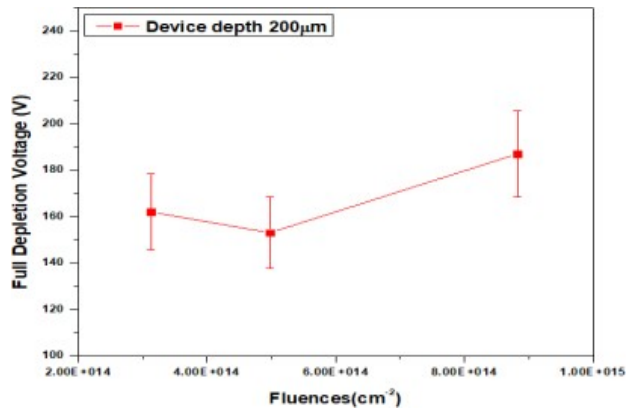

**Figure 3.** $V_{fd}$ as a function of mixed irradiated fluences in 200 μm p-MCz Si pad detector.

Using Mixed irradiation "four level deep traps model" for p-MCz Si PAD detector, the full depletion voltage $V_{fd}$ is shown as a function of equivalent fluence (Equivalent 1 MeV neutrons) in Figure 2(a). In figure 2(a), it is shown that the damage is accumulated at higher mixed fluences (>4.96 x$10^{14}$cm$^{-2}$) in p-MCz Si PAD and thus the $V_{fd}$ increases for both thick and thin p-MCz Si PAD detector (see figure 3).

The generation leakage current is not a major issue in mixed irradiated p-MCz Si PAD detectors even if, it increases the overall noise of the Si detector system. The increase of leakage current can be controlled using multiple field guard rings structure on outer surface of the detector (strip) and by cooling the detector system up to -20$^0$C to -30$^0$C in the new upgrade of the CMS experiment at HL-LHC. In the HEP experiment [5], temperature was measured a few cm away from the detector sample, which was stable within ±1˚C that result in 15-20% error in leakage current measurement as observed in figure 2 (b). In our previous work, a very good agreement in the experimental and SRH leakage current is shown at 297K in mixed irradiated n-MCz Si PAD detector [6], which was in good agreement as per our expectations.

It is noted that the $V_{fd}$ for thin detector is almost 50% less (< 200V) as compare to $V_{fd}$ of 300 μm thick p-MCz Si PAD detector for the same equivalent fluences (see figure 3), which is acceptable range as per the specifications of the irradiated detectors for the new CMS tracker detector at HL-LHC.

### 2.1.2 Introduction rate of E (30K) in mixed irradiated environment of n in p-MCz Si PAD detector

The introduction rate of E (30K) was not experimentally recorded accurately [7], which was due to the fluctuation in the TSC signal during measurements in the mixed irradiated detector. In order to understand the $V_{fd}$ behavior in p-MCz Si PAD detector, the introduction rate of E(30K) is studied deeply. It has been found that the introduction rate of E(30K) is not constant as per other donor / acceptor deep traps in mixed irradiated p-MCz Si PAD detector. The introduction rate of E(30K) decreases slightly with an increase in the fluences, which is shown in figure 4, and it is used as a fitting parameter for the good comparison of the experimental data and SRH data. Therefore, it cannot compensate the H (152K) acceptor trap. As a result in the p-MCz Si PAD detector, the damage is not accumulated at higher mixed fluences and thus, Vfd increases after (>4.96 x10$^{14}$cm$^{-2}$). It has been noticed that the effective introduction rate of E (30K) is higher in p-MCz Si than in mixed irradiated n-MCz Si PAD detector [6]. Whereas at high mixed fluences of 8.82 x10$^{14}$ n$_{eq.}$/cm$^2$ , it starts to saturate.

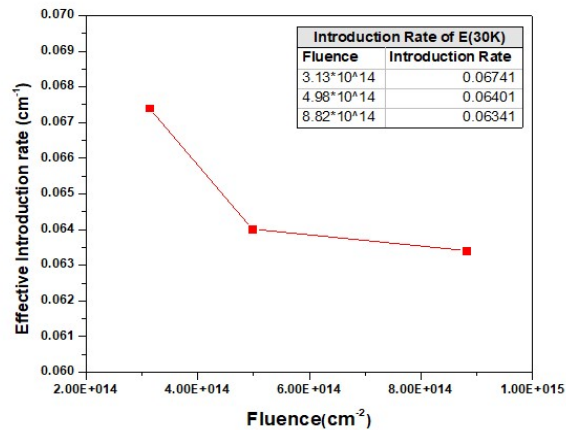

**Figure4.** Effective introduction rate of E(30K) as a function of mixed irradiation fluence.

## 3. Results and Discussion

In this section, the full depletion voltage and leakage current at higher mixed irradiation fluences are extrapolated in mixed irradiated thick and thin p-MCz Si PAD detector.

### 3.1 Extrapolated value of $V_{fd}$ in higher mixed irradiation environment

The extrapolated $V_{fd}$ value at higher mixed irradiation fluences for the thick and thin p-MCz Si PAD detector is shown in figure 5 (a). Due to the safety margin of the detector operation in the hostile radiation environment of the HL-LHC during the irradiation of detector, higher fluences of the order 4.50x10$^{15}$cm$^2$ are considered here for analysis. The extrapolated $V_{fd}$ value with 10% uncertainty as a function of the mixed irradiation fluence are plotted for the thick and thin p-MCz Si pad detector in figure5(a).

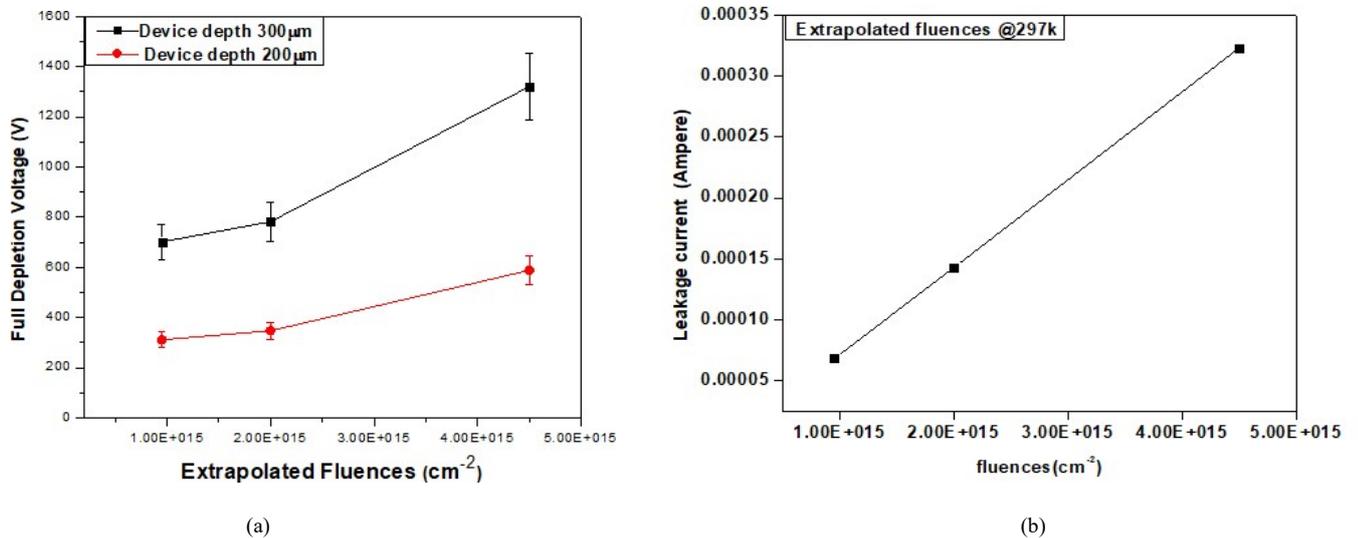

**Figure5.**(a) Extrapolated value of $V_{fd}$ for thick (300μm) and thin (200μm) in mixed irradiated p-MCz Si detector.(b) Leakage current as a function of Extrapolated mixed irradiated fluences at 297K after beneficial annealing of 80 min 60 degree.

The extrapolated value of $V_{fd}$ is nearly 1300 V at a fluence 4.5 x $10^{15}$ for the 300 μm mixed irradiated detector. Whereas in 200μm thin mixed irradiated p-MCz Si PAD detector, the $V_{fd}$ is around 500 V. Figure 5(b), shows the leakage current as a function of the mixed irradiated fluences at 297 K in thin detector. It has been observed that the leakage current increases linearly as expected in bulk damage Si detectors.

Finally, a four level deep trap mixed irradiation model for the p-MCz is proposed for the p-Mcz Si radiation damage analysis and the model can be used to design and optimize the radiation hard (mixed) p-MCz thin Si strip detector for new CMS tracker detector at HL-LHC experiments.

## 4. Conclusion

In this paper, a four level deep-trap mixed irradiation damage model for the p-MCz Si is proposed. A very good agreement has been observed in the experimental data and SRH / TCAD device simulation on the $V_{fd}$ of the mixed irradiated p-MCz Si microstrip detector. The uncertainty (within 10%) in leakage current is also in good agreement to the experimental observations. The extracted value of $V_{fd}$ of 200μm thin p-MCz Si strip detector irradiated by mixed irradiation fluence of 4.5x$10^{15}$cm$^{-2}$´Is less than 500V. Therefore, the radiation hard thin p- MCz Si microstrip detector for the new CMS tracker detector system can be designed and optimized for the HL-LHC experiments.